\documentclass[epj]{svjour}
\usepackage{graphics}
\begin{document}
\title{Anisotropic XY antiferromagnets in a field}
\author{W. Selke and S. Wessel}
\institute{Institute for Theoretical Solid State Physics, RWTH Aachen
  University, 52056 Aachen, Germany}
\date{Received: date / Revised version: date}
%
\abstract{
Classical anisotropic XY antiferromagnets in a field on 
square and simple cubic lattices are
studied using mainly Monte Carlo simulations. While in two dimensions
the ordered antiferromagnetic and spin--flop phases are observed to
be separated by
a narrow disordered phase, a line of direct transitions of first order
between the two phases and a bicritical point are found
in three dimensions. Results are compared to previous findings.
} 
\maketitle
\section{Introduction}
\label{intro}

Uniaxially anisotropic antiferromagnets in a
magnetic field have been studied quite extensively in the
past, both experimentally
and theoretically. Typically, they display, at low
temperatures, the antiferromagnetic (AF) phase and, when increasing
the field, the spin--flop (SF) phase \cite{Neel,Gorter}. In addition, more
complicated structures, like biconical \cite{KNF} ones, have been
observed. The various ordered phases may 
lead to interesting multicritical
behavior, including bi-- and tetracritical points \cite{KNF,LF,FN,Aha2}.

Experimentally, several antiferromagnets with uniaxial anisotropy 
have been investigated, three--dimensional magnets
\cite{SF,RG,FK,OU,FPB}, such as  MnF$_2$ and GdAlO$_3$, as
well as  quasi two--dimensional magnets \cite{SCD,T,BS,BSL}.

Much of the theoretical work is based on analyzing the prototypical 
classical Heisenberg model with uniaxial anisotropy, the XXZ
model, in a field, plus, possibly, further anisotropy
terms, such as single--ion anisotropies. Especially, mean--field
approximation \cite{Gorter,LF}, Monte Carlo
simulations \cite{LanBin,Holt1,hws,Zhou2,hs,sel,htl,fp}, and renormalization
group calculations \cite{KNF,FN,CPV,Folk,Eich} have been applied. 

In this article, we shall deal with the anisotropic
XY antiferromagnet in a field. The $x$--axis is taken to be the
easy axis and the field acts on the $x$--component
of the spins. The model is a variant of the
much studied uniaxially anisotropic three--component XXZ antiferromagnet
in a field, with the field along the easy axis, the $z$--direction. In a
previous paper, the two--dimensional version, on the square lattice, of
the XY model had been studied, applying ground state considerations
and Monte Carlo techniques \cite{hs}. At the field
separating the AF and SF phases, the
ground state has been found to be highly degenerate due to the
presence of biconical (or, more precisely, non--collinear,
biangular or bidirectional, BD) structures. At finite
temperatures, $T >0$, these structures seem to
lead to a disordered phase between the AF and SF phases, similar
to the situation in the two--dimensional XXZ Heisenberg
antiferromagnet. Here, we shall briefly
reconsider this case. Our main
emphasis will be on the anisotropic XY model on the simple cubic 
lattice. To our knowledge, no prior analysis exists. In particular, the
existence and nature of the possible bicritical point, at which
the AF, SF, and paramagnetic phases are expected to merge, in analogy to
the XXZ antiferromagnet, will be studied.

The paper is organised as follows: In the next section, the
anisotropic XY model will be introduced and ground state properties will
be discussed. Then, results for the model on the square lattice will
be presented, followed by our large--scale simulation
findings for the three-dimensional case. A short summary concludes
the article.

\section{The model}
The anisotropic XY antiferromagnet in a field is described by the
Hamiltonian

\begin{equation}
{\cal H}_{\mathrm{XY}} = J \sum\limits_{i,j} \left[ \, S_i^x S_j^x + \Delta S_i^y S_j^y \, \right] \; - \; H \sum\limits_{i} S_i^x
\end{equation}

\noindent
where the first sum runs over all pairs of neighbouring
sites, $i$ and $j$, of the lattice, with the second sum running
over all lattice sites. $J > 0$ is the antiferromagnetic coupling
constant, $\Delta <1$ is the anisotropy parameter, and $H$ is
the external field along the easy
axis, the $x$-axis. $S_i^{\alpha}$, $\alpha$=$x$ and $y$,
are the two components of the classical spin vector of length one.

\begin{figure}
\resizebox{0.95\columnwidth}{!}{%
  \includegraphics{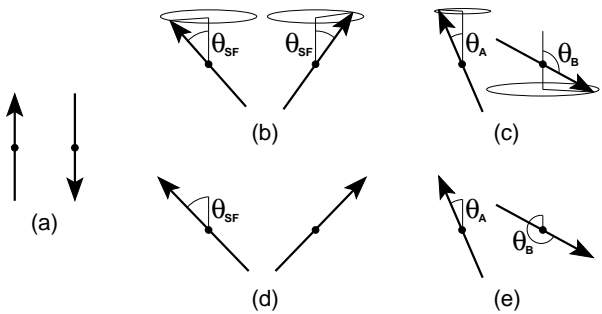}
}
\caption{Ground state configurations: (a) antiferromagnetic (AF),
 (b,d) spin--flop (SF), and (c) biconical as well as (e)
 bidirectional (BD) configurations. (a,d,e) occur in the anisotropic
 XY, (a,b,c) in the uniaxially anisotropic XXZ antiferromagnet. The 
 tilt angles are defined with respect to
 the easy axis, being the $x$--axis in the XY and the $z$--axis in
 the XXZ case.}
\label{fig:1}
\end{figure}

We shall consider mainly the model on a simple cubic lattice ($d=3$), but
some results for the square lattice ($d=2$) will also be discussed
comparing them to previous findings \cite{hs}. In both cases, the
ground states, at zero temperature, $T=0$, see Figure 1, may be
determined in a straightforward way \cite{LF,hs,TM}. For small
fields, $H < H_{c1}$, the AF
structure is stable, in which neighbouring sites belong to 
different sublattices, $A$ and $B$, where the $x$--components
of $A$-- and $B$--spins point in opposite directions. The $y$--component
vanishes at all sites. At intermediate
fields, $H_{c1} <H< H_{c2}$, one encounters
the SF structure, where the $x$--components of the spins on
both sublattices are equal, being smaller than one, while the
antiferromagnetic sublattice
structure shows up in the $y$--components.  At large
fields, $H > H_{c2}$, all spins
are aligned parallel to the field, with vanishing $y$--component
of the spin vectors. At $H=H_{c1}$, the
AF, the SF, and BD configurations form ground states. In
the highly degenerate BD configurations, the spins on the two 
sublattices, $A$ and $B$, are tilted with respect to the field
direction, the $x$--axis. The resulting two tilt angles, say, $\Theta_A$ and
$\Theta_B$, are interrelated interpolating
continuously between the AF and SF structures. The exact relation
depends on the anisotropy parameter $\Delta$. Of
course, in the SF limit, one
has $\Theta_A= \Theta_B= \Theta_{SF}(H) >0$.       

At non-zero temperatures, one expects, among others, phase
transitions in the Ising universality class from the 
disordered phase to the AF and SF phases. The 
longitudinal staggered magnetization, describing the antiferromagnetic
ordering of the $x$--component of the spins on the
$A$-- and $B$--sublattices, is
the order parameter in the AF phase. The transversal
staggered magnetization describes the
antiferromagnetic ordering of the $y$--component of the
spins in the SF phase, in which the $x$--component has the same value on
both sublattices. Accordingly, a possible bicritical
point is expected to belong to the $n=2$ or XY universality
class \cite{KNF,FN}. This behaviour
is in marked contrast to 
the situation in the anisotropic XXZ antiferromagnet in a
field, along the $z$--axis. There
the SF phase is described by an antiferromagnetic
ordering of the two components perpendicular to the
$z$--component of the spins, implying a transition to the
paramagnetic phase
in the XY universality class. Of course, the transition from
the paramagnetic to the AF phase belongs to the Ising
universality class, in the XXZ case as well. Thence, a 
possible bicritical point would fall into the $n=3$ or
Heisenberg universality class \cite{KNF,FN}.

To study the phase diagram, in the $(k_BT/J,H/J)$ plane, of the 
2d and 3d XY antiferromagnets, we did extensive Monte Carlo (MC) 
simulations, using the standard Metropolis algorithm \cite{LB}. Lattices
with $L^d$ sites were considered, with $L$ ranging from 10 to 200 for
$d=2$, and from 8 to 40 for $d=3$. In all cases, full periodic
boundary conditions were employed. As usual, finite--size extrapolations
were done to obtain estimates for the thermodynamic limit. Typically, runs
of, at least, $10^7$ MC steps per site (MCS) were performed, averaging 
over a few realizations, by using different random numbers, to estimate
error bars. Here, error bars are usually smaller than the size of
the symbols shown in the figures. Of course, close to the
phase transition, larger lattices may
require longer runs to take into account critical fluctuations and
critical slowing down \cite{LB}.

We recorded, among others, quantities related to
the longitudinal staggered magnetisation $M_{st}^x$
defined by $M_{st}^x= (S_A^x-S_B^x)/2$, and the corresponding
transversal staggered magnetization for the $y$ component of the 
spins. In particular, thermal averages over the second moments
of the magnetizations are expected to signal the transitions to
the AF and SF phases. Actually, these transitions
may be detected quite easily and reliably by the
Binder cumulant \cite{Binder}

\begin{equation}
  U^{x,y} = 1 - <(M_{st}^{x,y})^4>/(3 <(M_{st}^{x,y})^2>)
\end{equation}

\noindent
where the brackets denote thermal averages. The transition 
from the disordered phase to the AF phase can be determined by
$U^x$, and that to the SF phase can be determined
by $U^y$. To detect a possible bicritical point in
the $n=2$ or XY universality class, we also computed the
analoguous Binder cumulant $U^{xy}$, invoking the second and
fourth moments of the total staggered magnetization 
$M_{st}= \sqrt{((M_{st}^x)^2 + (M_{st}^y)^2)}$.

We also recorded the longitudinal and transversal
staggered susceptibilties, obtained, as usual, from the fluctuations
of the corresponding magnetizations. Further interesting information on
the thermal behavior of the model follows from the specific
heat, $C$, determined from the energy fluctuations, as 
well as histograms. In particular, we computed probability functions
of the tilt angles, such as the probability $p(\Theta)$ for encountering
the tilt angle $\Theta$ at an arbitrary site and the
probability $p_2(\Theta_A,\Theta_B)$ for finding the two
angles, $\Theta_A$ and $\Theta_B$, at
neighbouring sites of the lattice, as before \cite{hs}.

\section{Phase diagrams}

We shall first briefly discuss previous and new MC findings for the 2d
case, presenting evidence for a narrow disordered phase separating
the AF and SF phases. The main emphasis will be on the 3d model, for
which a qualitatively different topology of the phase diagram is
observed. In fact, our simulation data suggest the existence
of a bicritical point, at which the AF, SF, and paramagnetic phases
meet. In both cases, we set the anisotropy parameter $\Delta = 0.8$. 

\subsection{Anisotropic XY antiferromagnet on a square lattice} 

The phase diagram of the anisotropic XY antiferromagnet in a field on
a square lattice has been determined before \cite{hs}, comprising
the AF, SF, and paramagnetic phases. It is depicted in Figure
2, setting the anisotropy parameter $\Delta$ equal to 0.8. The
characteristic fields for the ground states are $H_{c1}/J = 2.4$ 
and $H_{c2}/J = 7.2$. The transition lines of the AF and SF phases
to the disordered phase, having confirmed to be in the
Ising universality class, approach each other rather closely at
$k_BT/J \approx 0.68$ and $H/J \approx 2.47$.

\begin{figure}
\resizebox{0.95\columnwidth}{!}{%
  \includegraphics{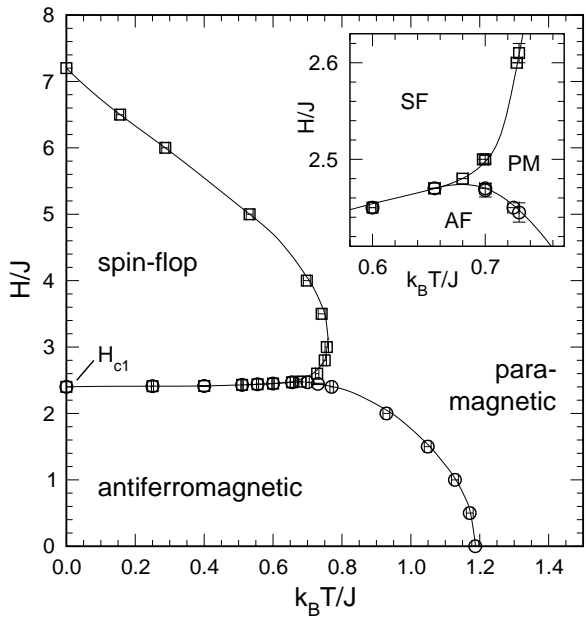}
}
\caption{Phase diagram of the anisotropic XY antiferromagnet 
   on the square lattice with $\Delta=0.8$. From Reference 20.}
\label{fig:2}
\end{figure}

In principle, the two lines might meet at a bicritical point in
the XY or Kosterlitz--Thouless \cite{KT} universality class, with
a line of transitions of first order between the AF and SF phases
at lower temperatures. Note that such a topology is excluded for the
two--dimensional XXZ antiferromagnet, where the bicritical
point would belong to the Heisenberg universality class. In   
fact, the existence of such a point at $T>0$ is
excluded by the well--known theorem of Mermin and Wagner \cite{MW}.

In our present study, we searched for possible evidence, whether
there is a bicritical point in the anisotropic XY model
on a square lattice. In particular, we did MC simulations at the
temperatures $k_BT/J$ = 0.7, 0.6, and 0.4.

\begin{figure}
\resizebox{0.95\columnwidth}{!}{%
  \includegraphics{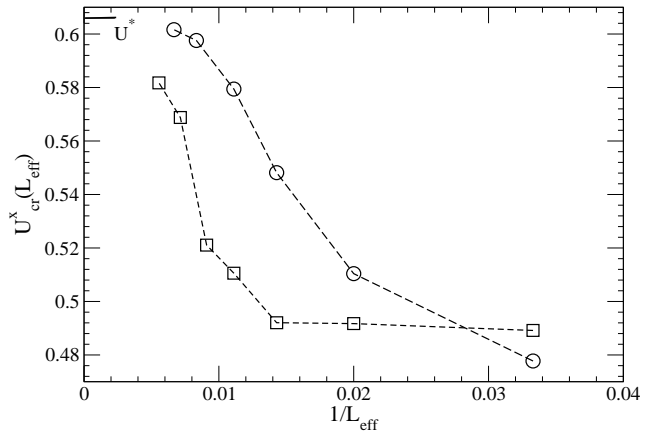}
}
\caption{Finite--size dependence of the crossing points
 of the longitudinal Binder cumulants $U_{cr}^x(L_{eff})$
 at $k_BT/J= 0.6$ (squares) and $0.7$ (circles). The value of the
 critical cumulant for the
 Ising case $U^*=0.61069...$ is indicated.}
\label{fig:3}
\end{figure}

The Binder cumulants $U^{x,y}$ turned out to be very useful in
investigating the low--temperature
region. We monitored the crossing points
$U_{cr}(L_{eff})$ of the
cumulants for two successive square lattices with linear dimensions
$L_1$ and $L_2$, assigning an effective length
$L_{eff}= (L_1+L_2)/2$. In general \cite{Binder}, one expects
$U_{cr}$ to occur in the thermodynamic limit at the phase
transition, with the critical value $U^*$. In the
case of phase transitions in
the 2d Ising universality class, employing full periodic
boundary conditions and considering systems with spatially 
isotropic spin--spin correlations \cite{cd,sesh}, one has
$U^*$= 0.61069 ..\cite{sesh,kabl}.

In Figure 3, results for the crossing points of the
longitudinal Binder cumulant $U^x$ at $k_BT/J$ = 0.6
and 0.7, with lattice sizes ranging from $L$= 20 to 200, are
depicted. The typical critical value, $U^*$= 0.61069..,
seems to be approached for both temperatures, with smaller 
finite--size effects at the higher 
temperature. Similar observations hold for the transversal
cumulant. Accordingly, the two transitions are seen to
be Ising--like. Note that at $k_BT/J$= 0.6, the two
distinct transitions occur at about $H/J$ = 2.4505
and 2.4525, as estimated from
the size--dependence of the crossing points, $U_{cr}^x$ and
$U_{cr}^y$. The disordered phase between the AF and SF
phases becomes more and more narrow, as the temperature
is lowered, see also Figure 2.

The existence of that narrow disordered phase had been
previously \cite{hs} inferred from histograms of the
probability $p_2$. $p_2$ describes the
relation between the tilt angles at neighbouring
sites, as stated above. Actually, at the temperature $k_BT/J = 0.4$, in an
extremely small range of fields, the
dominant configurations have neither an AF nor SF character, but
they are of BD type. The disordered phase had been argued to 
be due to BD fluctuations caused by the highly degenerate
ground state at $H_{c1}$. 

Indeed, the present simulations provide additional evidence
against a transition of first order between the AF and
SF phases at such low temperatures. For example, the
longitudinal and transversal Binder cumulants are expected to
become strongly negative near a first--order
transition \cite{BiLa}, even diverging to minus infinity in the
thermodynamic limit. We do not observe such a behavior. 

Moreover, we studied the maximal specific heat $C_{max}$, close
to the AF--SF transition as a function of the lattice
size, $L$. At a first order transition in two
dimensions,  $C_{max}$ is predicted \cite{FiBe} to grow, for 
sufficiently large lattices, like $C_{max} \propto L^2$. On
the other hand, at
an Ising--like transition in two dimensions, there is only
a logarithmic increase. Actually, we
fitted the simulation data for $C_{max}(L)$ to a
power law, $C_{max} \propto L^X$. We determined the
effective exponent $X_{eff}(L) = d \ln C_{max}/ d \ln L$. Simulations, 
at $k_BT/J$ = 0.4 and 0.6, are performed at discrete
values of $L$, ranging from 20 to 200. Data for successive
sizes, $L_1$ and $L_2$, are fitted to
the discretized effective exponent $X_{eff}(L_e)$, with
$L_e = \sqrt{(L_1\times L_2)}$. We find rather small
effective exponents, being, at most, about 0.25, far from
the quadratic behavior characterising a transition of first--order. 

In conclusion, for the anisotropic XY antiferromagnet on a square
lattice, we find no evidence for a direct transition of first
order between the AF and SF phases down to $k_B/J = 0.4$. Accordingly, there
is no indication for a bicritical point.

\subsection{Anisotropic XY antiferromagnet on a simple cubic lattice}

The crucial part of the phase diagram for the 3d
anisotropic XY model in a field is shown in Figure 4. It summarizes
present MC findings, setting the anisotropy parameter $\Delta$
equal to 0.8. At zero temperature, the AF configuration is stable up to
$H_{c1}/J = 3.6$ and the SF structure is stable up to
$H_{c2}/J = 10.8$.

\begin{figure}
\resizebox{0.95\columnwidth}{!}{%
  \includegraphics{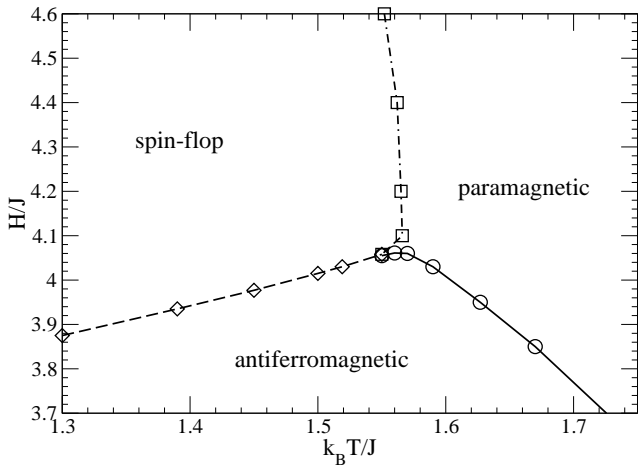}
}
\caption{Phase diagram of the XY antiferromagnet 
  with $\Delta=0.8$ on the simple cubic lattice.}
\label{fig:4}
\end{figure}

The transition lines between the disordered paramagnetic
phase and the two orderded phases AF and SF phases are
believed to belong
to the Ising universality class. In fact, our estimates
for the critical exponents  of the staggered magnetizations and 
susceptibilities as well as of the specific heat agree with
the Ising values \cite{Vic}. Furthermore, for both
transitions, the critical Binder
cumulants, $U^*$, are found to be close to the
characteristic value for isotropic phase transitions
of 3d Ising type, $U^* \approx 0.465$ \cite{Hb}.

\begin{figure}
\resizebox{0.95\columnwidth}{!}{%
  \includegraphics{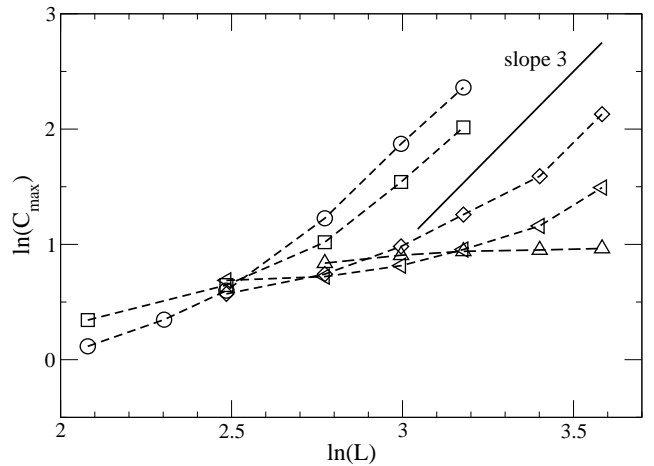}
}
\caption{Maximal specific heat, $C_{max}$, versus lattice size, $L$, near the 
 AF--SF transition at $k_BT/J$ = 1.0 (circles), 1.3 (squares), 
 1.45 (diamonds), 1.50 (triangles left), and 1.55 (triangles up). The
 straight line corresponds to $C_{max} \propto L^3$.}
\label{fig:5}
\end{figure}

The most interesting aspect of the phase diagram concerns
the existence, location, and
characteristics of the bicritical point, at which the AF, SF, and
paramagnetic phases may meet. 

To identify a, possibly, first--order transition between the AF and
SF phases at low temperatures, we monitored the size--dependence
of the maximum in the specific heat, $C_{max}$. A first
transition of first order is expected \cite{FiBe} to be signalled by a
divergence of the form $C_{max} \propto L^3$, for sufficiently large
lattice sizes. In contrast, at a continuous transition in three
dimensions, one expects a power--law
behavior with an exponent $\alpha/\nu$, with
$\alpha$ and $\nu$ being the standard critical
exponents. For an Ising--type 
transition in 3d \cite{Vic}, one obtains
$\alpha/\nu \approx 0.175$. As depicted in Figure 5, the
size--dependence typical for a first--order transition is observed to be
approximated closely at low temperatures. For instance, at 
$k_BT/J$ = 1.0, the cubic size dependence, characteristic
for a first--order transition, is
approached already for $L$ = 16. Increasing the
temperature, the crossover towards the (almost) cubic behaviour is
shifted to larger lattices, being about $L$ = 30 at
$k_BT/J$ = 1.45. Obviously, at
higher temperatures, as shown in Figure 5 for $k_BT/J = 1.5$, it becomes
more and more demanding to reach the asymptotic regime. Presumably, the
difficulty is related to
the increase in the correlation length at the
first--order transition when getting closer to the
bicritical point. Actually, at the bicritical point in
the XY universality class, one expects no divergence in
$C$, but a cusp--like singularity, with the corresponding
standard critical exponent $\alpha$ being slightly
negative \cite{Vic}. Indeed, see Figure 5, at
$k_BT/J = 1.55$, $C_{max}$ increases only slightly 
with the lattice size, indicating, probably, the closeness of the
bicritical point. 

Another signature of the first--order transition at low temperatures 
is provided by the longitudinal and transversal Binder
cumulants, $U^x$ and $U^y$. In fact, both quantities display, close
to the boundary between the AF and SF phases, negative
minima, choosing temperatures $k_BT/J$ ranging 
from 1.0 to 1.50. Lattices with $L$ up to 36
were considered. The minima became more pronounced, when
lowering the temperature, reflecting the fact that then the transition
gets more strongly of first order, with a smaller
correlation length, in agreement with the behavior of the
specific heat, as discussed above. By increasing the lattice
size, the minima are getting deeper. In fact, in the
thermodynamic limit, the cumulant
is predicted to be minus infinity at the phase 
transition of first order \cite{BiLa}.
   
\begin{figure}
\resizebox{0.95\columnwidth}{!}{%
  \includegraphics{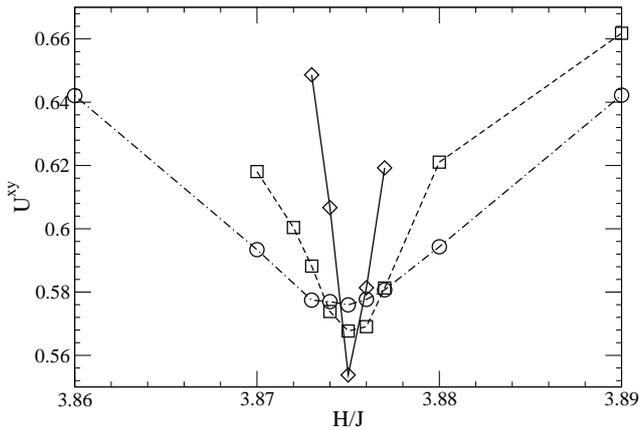}
}
\caption{Binder cumulant $U^{xy}$ at constant temperature, $k_BT/J$ = 1.3,
  as a function of the field $H/J$ near the AF--SF transition for
  lattices with $L$ =16 (circles), 20 (squares) and 30 (diamonds).}
\label{fig:6}
\end{figure}

The data on the specific heat $C$ suggest that the transition between
the AF and SF phases is of first order up to temperatures of, at least,
about $k_BT/J \approx 1.5$.  To locate the bicrital point 
and to characterize its nature, the Binder cumulant
$U^{xy}$ turns out to be quite useful as well. Its
critical value, at a
transition point in the XY--universality class in
three dimensions, has
been estimated \cite{HT} to be $U^{(xy)*} \approx 0.586$.

Of course, $U^{xy}$ is expected to approach, for large
lattices, in both ordered phases the limiting value 
2/3. When the transition between the AF and SF
phases is of first order, as
suggested by the specific heat, the cumulant
displays, fixing the temperature
and varying the field (or vice versa), a minimum close to
the transition, $U^{xy}_{min}$, as exemplified in Figure 6. When
increasing the lattice size, the minimum is, eventually, 
lowered, falling clearly below the value, 0.586, characterising
the XY universality class. Results of the present simulations
on the size--dependence of the height of the minimum for various 
temperatures and fields are depicted in Figure 7.
 
\begin{figure}
\resizebox{0.95\columnwidth}{!}{%
  \includegraphics{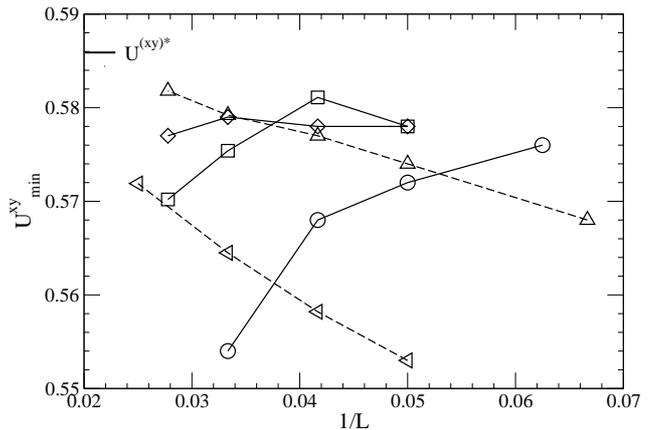}
}
\caption{Minimum of the xy--cumulant $U^{xy}_{min}$ as a function
  of lattice size $L$ for various temperatures $k_BT/J$ = 1.3 (circles),
  1.45 (squares), 1.5 (diamonds), and 1.55 (triangles left) as well as for
  fixed field $H/J$ = 4.03 (triangles up). The critical cumulant in
  the XY universality class occurs at $U^{(xy)*} \approx 0.586$ \cite{HT}.}
\label{fig:7}
\end{figure}

Perhaps most interestingly, Figure 7 shows a rather drastic
change in the size dependence of the height of the minimum, when
getting closer to the bicritical point. Eventually, $U^{xy}_{min}$ starts
to increase for large lattices. It seems to approach, for
sufficiently large lattices, the 
characteristic value of the 3d XY universality
class, $U^{(xy)*} \approx 0.586$, in the  
temperature range between about $k_BT/J$ = 1.51 and 1.55. The
lower bound is inferred from the simulation data when fixing
the field at $H/J$ = 4.03, with
the corresponding phase transition occurring at
$k_BT/J \approx 1.518$. Accordingly, the present
simulations are consistent with the existence of a bicritical
point of XY type in the three--dimensional model. The
result confirms the previous theoretical prediction \cite{KNF,FN}
on the nature of the bicritical point, based
on renormalization group calculations. Note that at even
higher temperatures, either the two transition lines of the AF and SF
phases to the paramagnetic phase are well separated or
there is only the transition to the AF phase, as depicted
in Figure 3. Then, the transition is no longer signalled by a
minimum in $U^{xy}$.

Of course, to explore, in detail, finite--size
effects in the crossover region in the vicinity of
the bicritical point, data of very high accuracy for larger
lattices would be desirable. This feature, however, is beyond
the scope of our study.

\section{Summary}

We conclude that the topology of the phase diagram, in
the ($k_BT/J, H/J$) plane, of the anisotropic XY antiferromagnet
depends significantly on the lattice dimension. For the square
lattice, in agreement with a previous study, we find no evidence
for a bicritical point. Instead, in between the AF
and SF phases, there is a narrow intervening paramagnetic phase down
to the lowest temperatures we considered. In contrast, for the simple
cubic lattice, we find a line of direct transitions of
first order between the two ordered phases, leading, eventually, to
a bicritical point.

We locate the bicritical point
at $k_BT/J \approx 1.53 \pm 0.02$, based on finite--size analyzes
for the specific heat and Binder cumulants. The point belongs to
the XY universality class. 

The qualitatively different phase diagrams in
two and three dimensions may be explained
by the bidirectional configurations, leading
to the highly degenerate ground state at the field separating the
AF and SF structures. In two dimensions, these configurations
seem to surpress the direct transition between the AF and SF phases, while
they are thermally less relevant in three dimensions. Such a 
feature has been seen to hold in the related XXZ Heisenberg antiferromagnet
in a field as well. In that case, the existence of a bicritical
point (belonging to the Heisenberg universality class) in two dimensions
is already excluded by the Mermin--Wagner theorem. 

We should like to thank especially Kurt Binder, David Landau, 
Reinhard Folk, and Martin Holtschneider for useful information, remarks,
and discussions on the topic of this article. The article is dedicated
to Wolfhard Janke on the occasion of his sixtieth
birthday. We thank him for very helpful and pleasant
interactions.

\end{document}